\documentclass{webofc}
\usepackage[varg]{txfonts}   % Web of Conferences font
\usepackage{siunitx}
\usepackage{subfig} 
\usepackage{threeparttable}
\woctitle{RICAP-14 The Roma International Conference On Astroparticle Physics}
\begin{document}
\title{Analysis of the cumulative neutrino flux from Fermi-LAT blazar populations using 3 years of IceCube data}

\author{Thorsten Gl\"usenkamp\inst{1}\fnsep\thanks{\email{thorsten.gluesenkamp@desy.de}} for the IceCube Collaboration
        % etc.
}

\institute{DESY Zeuthen, Platanenallee 6, 15738 Zeuthen, Germany 
          }

\abstract{
The recent discovery of a diffuse neutrino flux up to PeV energies raises the question of which populations of astrophysical sources contribute to this diffuse signal. One extragalactic candidate source population to produce high-energy neutrinos are Blazars. We present results from a likelihood analysis searching for cumulative neutrino emission from Blazar populations selected with the 2nd Fermi-LAT AGN catalog (2LAC) using an IceCube data set that has been optimized for the detection of individual sources. In contrast to previous searches with IceCube, the investigated populations contain up to hundreds of sources, the biggest one being the entire Blazar sample measured by the Fermi-LAT. No significant neutrino signal was found from any of these populations. Some implications of this non-observation for the origin of the observed PeV diffuse signal will be discussed.
%Beamed jet-on radio-loud AGN, Blazars, are candidates for (U)HECRs and thus for high energy neutrinos.Since the discovery of the diffuse TeV-PeV neutrino signal it has been unclear if and how much Blazars contribute to this flux. This talk investigates this question.We use the 2nd-year FERMI-LAT AGN catalog to define five GeV-Blazar populations,
%the biggest one being all 862 Blazars together, and perform a neutrino correlation study %using 3 years of IceCube muon-track data with each of those. 
%No significant signal is seen, which results in flux limits smaller than \SI{17}{\percent} of the sky-integrated diffuse signal recently reported by IceCube assuming a 1-1-1 composition of species and a spectral index of $-2.5$ similar to current best fit estimates of the diffuse flux. The analysis is done in such a way, that the limits do not require to assume a perfect correlation of neutrino emission with gamma ray emission and is also valid for quasi-isotropic sub-populations, e.g. TeV-emitting Blazars only. For FSRQ's, the limit can be extended from the FERMI-LAT FSRQ's to all FSRQ's in the observable universe, assuming a correlation between the gamma energy flux and the neutrino flux and exploiting the fact that FERMI resolves the majority of the GeV-gamma emission of FSRQ's. In this case the contribution of FSRQ's is found to be less than \SI{3}{\percent}.
}
\maketitle
%

%{\bf TODO: Call the "All Blazars" sample the "All 2LAC Blazars" sample}
\section{Introduction}
\label{intro}

%The IceCube Neutrino Observatory located at the geographic SouthPole is currently the largest high energy neutrino detector in the world. With over 5000 PMT's buried between \SI{2.5}{\kilo \metre} and \SI{1.5}{\kilo \metre} depth within the Antarctic ice, it covers a volume of roughly \SI{1}{\kilo \metre \cubed} (see figure \ref{}) to detect charged secondaries from neutrino interactions via Cherenkov Light. 
In 2013, the IceCube collaboration reported on the detection of an astrophysical diffuse neutrino flux in the \SI{40}{TeV} - \SI{2}{PeV} energy range \cite{Aartsen2013a}. Since then it has been measured down to about \SI{10}{TeV} in energy and the spectrum has been shown to be consistent with a power law of index $-2.5$ \cite{Aartsen2015}. The origin of the observed neutrino signal is so far unidentified. A prominent candidate is the population of Blazars, active galactic nuclei (AGN) with their relativistic jets pointing along the line-of-sight. Neutrino emission from Blazars has been discussed extensively in hadronic jet models originating from $p$-$p$ or $p$-$\gamma$ interactions (e.g. \cite{Mannheim1995}) which also produce gamma rays from neutral pion decays. %As such, they are a potential candidate to contribute to the diffuse neutrino flux, especially since 
The extragalactic non-thermal radiation background at GeV energies and above (EGB) is dominated by the emission from Blazars \cite{Inoue2014}. %Such non-thermal radiation is inevitably produced alongside the neutrinos in the decay of pions, and its presence might therefore be interpreted as a hint that Blazars are also the sources of the astrophysical neutrinos. However, the degree and shape of correlation between the neutrino $\nu$ and $\gamma$ emission spectra of Blazar jets is difficult to calculate as it depends on intrinsic source parameters, and the radiation environment between source and observer. Indeed, in most hadronic Blazar jet models, $p-\gamma$ interactions lead to peak neutrino energies above few PeV, while a rather soft spectrum is observed by IceCube between tens of TeV and few PeV. This has lead some theorists to conclude that they are likely subdominant sources for the diffuse IceCube signal, even the flat spectrum radio quasars (FSRQ's), a sub-class of Blazars, which is expected to have rather lower peak energies for the neutrino spectrum due to the presence external radatiation fields at these sources \cite{Dermer2014}.
An experimental approach is presented here to answer the question if the Blazars that dominate the high energy EGB emission also produce a large fraction of the astrophysical neutrino flux. The idea is to look for directional clustering in a large sample of muon tracks collected by IceCube, around the directions of gamma-ray sources associated with Blazars in the \textit{Fermi}-2LAC catalog\cite{Ackermann2011}. %The overwhelming majority of the muon tracks is of atmospheric origin. However, while the atmospheric background is isotropic, the few muon tracks from the interactions of astrophysical neutrinos cluster strongly around the source directions, as their directions can be typically reconstructed to sub-degree accuracy.

This analysis is performed as a maximum likelihood stacking analysis looking for a cumulative signal from multiple sources. In contrast to previous stacking searches which looked at 10-30 sources  \cite{Achterberg2006} \cite{Aartsen2014a}, the populations in this work usually comprise more than 100 sources, the biggest population being 862 sources when all Blazars from the 2LAC catalog are combined. Any search for a cumulative signal from multiple sources needs to make assumptions about the relative contributions (``weights'') of the individual sources within the population. Here, this weighting scheme is set up in a way to be minimally biased with respect to $\gamma-\nu$ correlation assumptions. Section \ref{sec-1} covers the motivation for this type of search and the technical setup. In Section \ref{sec-2} the sample of muon tracks used in this analysis is described. In Section \ref{sec-3} the results of this search are summarized. Conclusions are presented in Section \ref{sec-4}.

\section{Motivation and Method}
\label{sec-1}

In previous stacking searches for a cumulative neutrino signal from Blazars  in IceCube data \cite{Achterberg2006} \cite{Aartsen2014a}, the measured gamma-ray flux was taken as an estimate for the expected neutrino flux. However, the $\gamma$ and $\nu$ spectra may not be strongly correlated: Other contributions to the gamma flux almost surely exist, e.g. inverse Compton photons \cite{Bloom1996} from high-energy electrons or synchrotron photons from muons and protons \cite{Muecke2003}. Moreover, radiation fields at the sources and in the interstellar medium absorb and reprocess the gamma rays originally produced together with the neutrinos. In order to overcome this bias as much as possible, two modifications with respect to previous IceCube stacking searches have been implemented: 

\begin{itemize}

\item \textbf{There is no sub-selection of sources within the catalog based on their gamma-ray flux}

The populations tested for a neutrino signal are only defined via their spectral classification, the largest one being just all Blazars of the 2LAC catalog. Due to the competing production processes and different environmental conditions described above, a source with a relatively low gamma-ray flux could still contribute significantly to the neutrino signal. %with a high gamma flux might be leptonically dominated with a low neutrino output. As such, the gamma ray flux is only %used as a tracer to define the "search region", nothing more. 
% In weighting scheme 1 (see next section), which weights each source with the gamma ray energy flux, low gamma sources are anyway completely suppressed and the result is rather similar to a pre-cut on gamma-ray flux to reduce the catalogue size.

\item \textbf{Two complementary weighting schemes are used, including a uniform relative weight for all sources}
\end{itemize}

\begin{enumerate}

\item $w_{source} \propto F_{\gamma}$ 

Note that $F_\gamma$ is the energy flux. Gamma-rays produced in neutral pion decays at multi-TeV energies (i.e. by the same process that also produces a neutrino signal) are absorbed quickly by IR photons present in the source environment and reprocessed to lower energies. In this process they would approximately maintain their bolometric luminosity.

\item $w_{source} = 1$

Equal weighting is used as the most unbiased weighting scheme, completely ignoring potential correlations. Although it is in a strict sense unrealistic, because no population consist of sources which all emit equally strong, the resulting limits are quite independent of the unknown correlation between neutrino and gamma-ray flux. They even hold for the case that no correlation exists.

\end{enumerate}
The analysis is performed with an unbinned maximum likelihood Ansatz. The log-likelihood function is constructed as
\begin{equation}
\mathrm{ln}(L)=\sum_{i}^{N}{ \mathrm{ln} \left(\frac{n_s}{N} \cdot S_i + \left(1-\frac{n_s}{N} \right)\cdot B_i  \right)  }
\label{eq:llh}
\end{equation}
where N is the total number of neutrino events, $S_i$ and $B_i$ are the signal probability distribution function (PDF) and background PDF evaluations for event $i$. The background PDF is taken from the data distribution in declination and reconstructed energy and is uniform in right ascension. The signal PDF is defined as %a linear superpostion of point spread functions $S_j$ (PSF's) centered at the source positions reflecting a composite population.
\begin{equation}
S_i=\frac{\sum_{j}^{N_{\mathrm{sources}}}{w_{\mathrm{j,tot}} \cdot S_j(x_i;\sigma_i) \cdot \varepsilon_j(E_i)}}{\sum_{j}^{N_{\mathrm{sources}}}{w_{\mathrm{j,tot}}}}  , \ \ w_{tot}=w_{\mathrm{source}} \cdot w_{\mathrm{dec.}}
\end{equation}
where $j$ denotes the individual sources within the population. The term $S_j(x_i;\sigma_i)$ denotes the point spread function (PSF) of source $j$ evaluated at celestial coordinate $x_i$, depending on the event-specific angular error estimate $\sigma_i$, while the term $\varepsilon_j(E_i)$ denotes the energy PDF of source $j$ evaluated for the reconstructed energy $E_i$. The PSF is locally modeled as an analytic 2-d radially symmetric gaussian while the energy PDF $\varepsilon$ is estimated from Monte Carlo simulations. Each individual source within the signal PDF is weighted with a declination-dependent detector response weight $w_{\mathrm{dec.}}$ and the relative source weight $w_{\mathrm{source}}$ described above. 
Equation \ref{eq:llh} is minimized with respect to $n_s$ and the likelihood ratio with respect to $n_s=0$ is calculated for each population and each weighting scheme.
It is tested if the observed likelihood ratio is compatible with statistical fluctuations of the background events by repeating the process for many randomized sky maps. A randomized sky map is generated by randomizing the right ascension of the observed events (a more detailed description of this procedure is found in \cite{Aartsen2014a}). 
%This simple proced is possible due to the geographic position of the detector and the large number of events, while %signal is injected at source positions of a given population according to a given spectrum. 
%
%This is again done for the real sky and compared to the LLH-ratio of simulations. 
In case no significant deviations from background fluctuations are observed, upper limits on the neutrino flux of the tested population are calculated via the $CL_s$ method \cite{Olive2014}. This is performed twice, once for each weighting scheme.

\section{Data and source catalogs}
\label{sec-2}

\subsection{Muon Track Data}

%A stacking analysis looking for spacial correlations requires optimal directional reconstruction of the primary neutrino. Muon neutrinos result in muon tracks which can be reconstructed with a median angular resolution \SI{1}{\degree} at \SI{10}{TeV} of the primary neutrino direction ( ) and fulfill this requirement in IceCube.
A muon-track selection for the years 2009-2011 (IC59-IC86/1) is used, which is similar to the one described in \cite{Aartsen2014a}\footnote{The IC79 dataset is the reprocessed one also used in the Blazar stacking search in \cite{Aartsen2014a}.}. The median angular resolution of muon-neutrino tracks with this selection is around \SI{1}{\degree} at \SI{1}{\tera \eV} and \SI{0.5}{\degree} at \SI{100}{\tera \eV}. One additional cut is applied which keeps only events that have an estimated angular resolution better than $5$ degrees. In the southern hemisphere (downgoing events in IceCube) this sample is dominated by muons from air showers, while in the upgoing region the majority of events come from atmospheric muon-neutrinos. %The actual event number per year are given in table \ref{tab-event_no}.
%\begin{table}
%\centering
%\caption{Track-like events per year. Variations are due to different selections.}
%\label{tab-event_no}       % Give a unique label
% For LaTeX tables you can use
%\begin{tabular}{ll}
%\hline
%Year & no. of events  \\\hline
%2008/2009 & 107011 \\
%2009/2010 & 93720 \\
%2010/2011 & 136245  \\\hline
%\end{tabular}
% Or use
%\vspace*{5cm}  % with the correct table height
%\end{table}

\subsection{Populations tested}

The populations to be tested are based on the 2nd Fermi AGN catalog (2LAC) \cite{Ackermann2011} and are classified purely by spectral properties. Only sources that do not suffer from potential confusion are considered (CLEAN flag set to True). The populations are listed in table \ref{tab-populations}. It should be noted that some of the populations are overlapping and the p-values correlated (the largest overlap exists between the LSP and FSRQ populations, which is about \SI{60}{\percent}).
\begin{table}
\begin{threeparttable}
\centering
\caption{Definitions of Blazar populations}
\label{tab-populations}       % Give a unique label
% For LaTeX tables you can use
\begin{tabular}{lll}
\hline
Type & No. of sources & Motivation  \\\hline
All 2LAC Blazars & 862 & No bias \\
FSRQ\tnote{1} & 310 & BLR radiation \cite{Atoyan2001} \\
LSP\tnote{2} & 308 &  FSRQ and LSP-BLLAC might be intrinsically similar \cite{Giommi2012} \\
ISP | HSP\tnote{2} & 301 & HSP objects seem to evolve differently \cite{Ajello2013} \\
LSP \& BLLAC\tnote{1}\tnote{2} & 62 & Motivated by work in \cite{Muecke2003} \\\hline
\end{tabular}
\begin{tablenotes}
\item{1} FSRQ/BL-LAC based on optical line equivalent width
\item{2} Low/Intermediate/High Synchroton Peaked Object (LSP/ISP/HSP), based on position of synchroton peak
\end{tablenotes}
% Or use
%\vspace*{5cm}  % with the correct table height
\end{threeparttable}
\end{table}
%The LSP sample is motivated by the fact that LSP-BLLACs and FSRQs might be intrinsically similar. The ISP/HSP sample is motivated because HSP objects show different evolution (ISP objects are included to increase the sample size).
%LSP-BLLACs are motivated by the model in \cite{Muecke2003}
%The LSP and ISP|HSP population, the LSP-BLLAC and FSRQ population and the LSP-BLLAC and ISP|HSP populations are non-overlapping. The overlap between the FSRQ and ISP|HSP population is less than \SI{1}{\percent} while the overlap between the FSRQ and LSP population is around \SI{60}{\percent}.
%{\bf TODO: Comment on how independent the samples are.}
\section{Results} 
\label{sec-3}
\begin{figure}
\centering
\subfloat[]{\label{fig-1-a}\includegraphics[width=7cm]{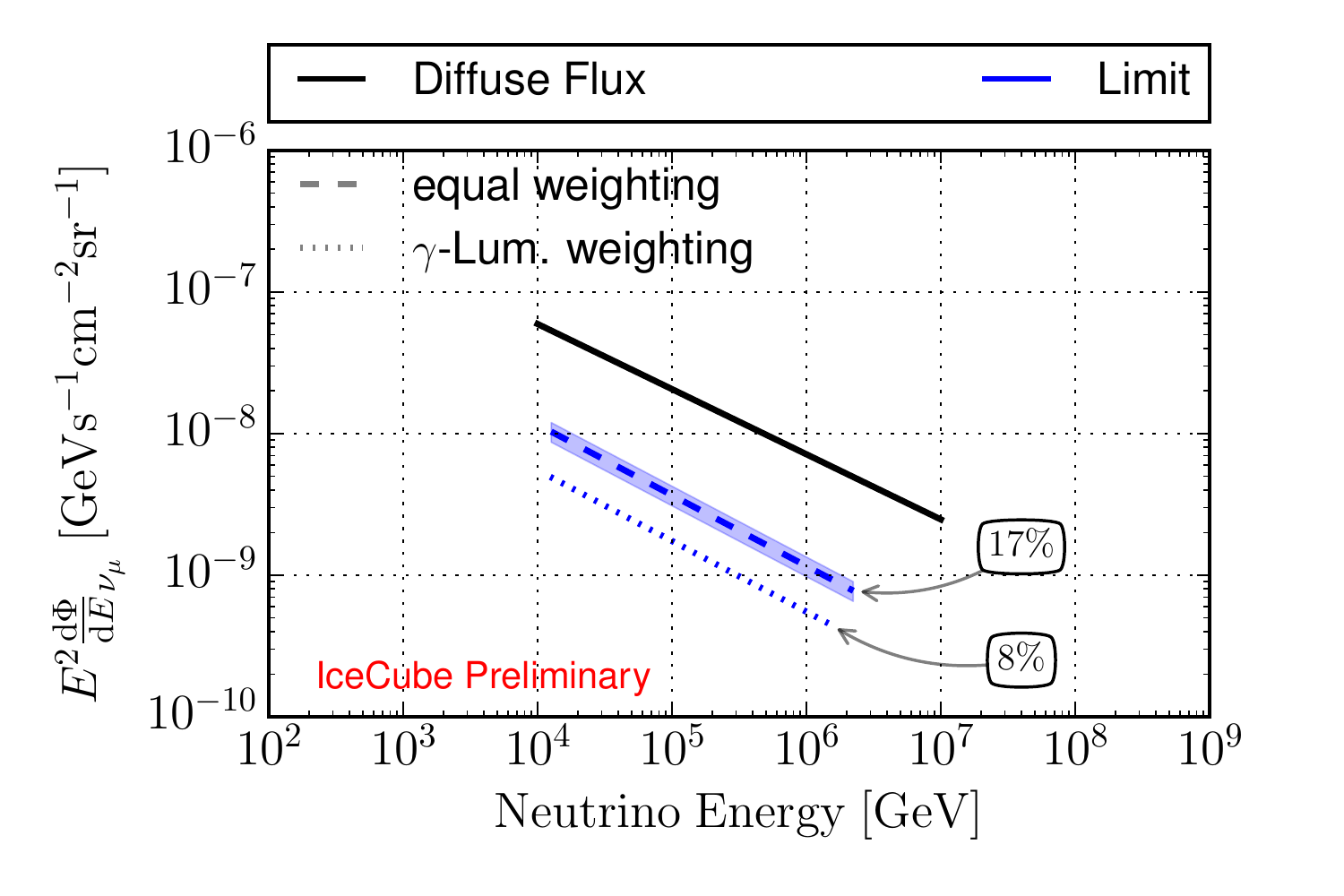}}
\subfloat[]{\label{fig-1-b}\includegraphics[width=7cm]{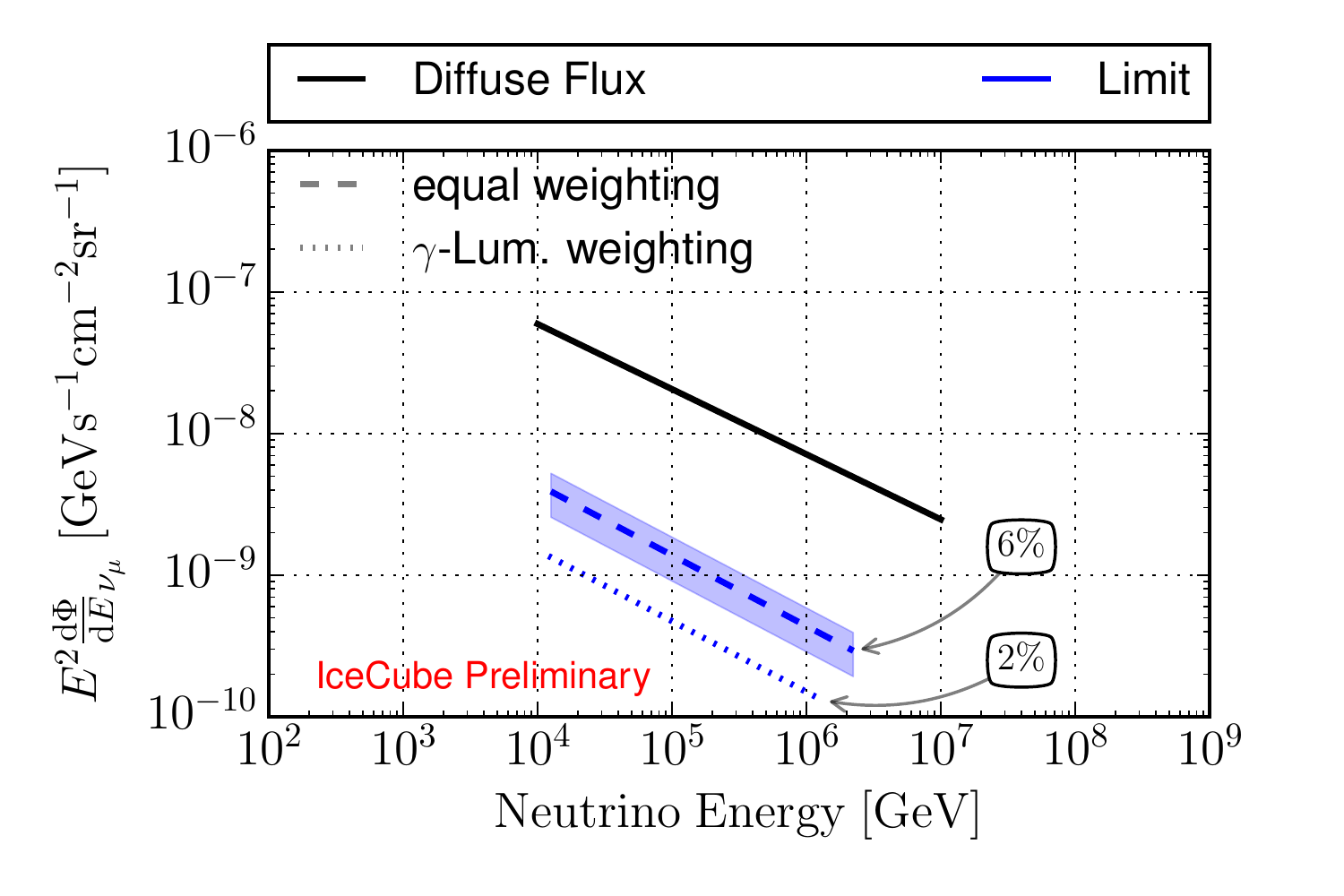}}
\caption{Neutrino flux upper limits for an $E^{-2.5}$ spectrum (blue) compared to diffuse bestfit (black solid) from \cite{Aartsen2015}. a) shows this comparison for the "All 2LAC Blazar" sample, b) for the FSRQs. Percentages with arrows denote the fraction with respect to the diffuse flux.}
\label{fig-1}       % Give a unique label
\end{figure}

For each population two tests were performed, one for each of the two different source weighting schemes. Table \ref{tab-results} summarizes the results\footnote{No trial-correction for testing multiple populations is included.}. All test outcomes are compatible with statistical fluctuations of the background. The smallest p-value is \SI{6}{\percent} for the "All 2LAC Blazar" sample when tested with the "equal weighting" scheme.
\begin{table}
\centering
\caption{Results of the Blazar population tests for both weighting schemes.}
\label{tab-results}       % Give a unique label
% For LaTeX tables you can use
\begin{tabular}{ccc}
\cline{2-3}
 & \multicolumn{2}{c}{p-values} \\
 & $w_{source} \propto F_{\gamma}$  & $w_{source} = 1$ \\\hline
 All 2LAC Blazars & \SI{36}{\percent} &\SI{6}{\percent} \\\hline
 FSRQs & \SI{34}{\percent} & \SI{34}{\percent} \\\hline
 LSPs & \SI{36}{\percent} & \SI{28}{\percent} \\\hline
 ISP/HSPs & >\SI{50}{\percent} & \SI{11}{\percent} \\\hline
 LSP-BLLACs & \SI{13}{\percent} & \SI{7}{\percent} \\\hline
\end{tabular}
% Or use
%\vspace*{5cm}  % with the correct table height
\end{table}
 %Their calculation would be non-trivial due to overlap of the populations and of the two weighting schemes.  
%Upper limits on the neutrino flux from the different Blazar populations have been calculated. These upper limits depend strongly on the assumptions about the energy spectrum of the neutrinos emitted from the Blazars. %Figure \ref{} shows the corresponding limits calculated for generic power-law spectra with indices $\Gamma=-2.0$ and $\Gamma=-2.7$ for the "All Blazar" population.
%For two-column wide figures use syntax of figure~\ref{fig-2}
%\begin{figure*}
%\centering
%\includegraphics[width=14cm,clip]{combined_generic.pdf}
%\vspace*{5cm}       % Give the correct figure height in cm
%\caption{Neutrino flux upper limits for the "All Blazar" population for two different power-law spectra with indices $\Gamma=-2.0$ (green) and $\Gamma=-2.7$ (red). The respective upper limit (dashed) shows the limit from weighting scheme 1 (uniform weighting scheme). The band around this limit encompasses \SI{90}{\percent} of random realizations of the differential blazar source count distribution (two realizations shown in orange and blue). The respective lower limit (dotted) shows the limit for weighting scheme 1 (energy flux weighting). The energy range in which the limit is plotted is the range which contributes \SI{90}{\percent} to the total sensitivity for the given spectral model. }
%\label{fig-generic}       % Give a unique label
%\end{figure*}
Figure \ref{fig-1} shows the resulting neutrino flux upper limits for an $E^{-2.5}$ spectrum for the "All 2LAC-Blazars" population and for the FSRQ population.
Notice that the flux upper limit using the equal-weighting scheme is represented as a band, instead of a line. It serves as an indicator of how robust this flux limit is towards changes of the actual source count distribution over the sky, which is unknown. Random realizations are created assuming a source count distribution for the neutrino emission that has the same shape as the source count distribution in gamma rays (parametrization from \cite{Abdo2010}). %Such a proxy distribution is needed since the real neutrino source count distribution is not known. 
For each source in the population a relative source strength is drawn from the source count distribution and assigned to the source.
%Then the neutrino signal is simulated with the obtained distribution of relative flux strengths, and the corresponding flux upper limit calculated.
The band encompasses the central \SI{90}{\percent} of the flux upper limits found from different such realizations. In comparison to the limits the best fit diffuse flux above 10 TeV is shown from \cite{Aartsen2015} with a spectral index of $-2.46$. Assuming an isotropic flux and 1/1/1 composition of flavors, the maximal contribution from the Blazar populations with respect to the diffuse flux can be read off. This contribution to the diffuse neutrino flux is at most around \SI{20}{\percent}, even if there is a very weak correlation with the observed gamma flux. For FSRQ's, these numbers are lower as the population size is smaller. %which is compatible with $-2.5$ within one standard deviation.

%Importantly, this limit also approximately holds for almost arbitrary sub-populations, as long as they are distributed %quasi-isotropic over the whole sky, such as the TeV-Blazars of the catalogue. This is because their intrinsic source %count distribution can be seen as just one random realization which is likely close to the equal limit band. This has %been verified in simulations.
%For the Energy flux weighting the injected neutrino strengths are assumed to follow the energy fluxes, and thus results in a single limit line. 

%{\bf TODO: Give table with limits for populations / could be concatenated with p-value table}
\section{Conclusion}
\label{sec-4}

In summary, the Blazars in the 2LAC catalog or any of the tested sub-samples are not responsible for the majority of the observed diffuse astrophysical neutrinos. This is in particular interesting since these Blazars dominate the high-energy extragalactic gamma-ray sky. The result is also valid if only a weak neutrino flux correlation with the observed gamma ray emission is assumed and expected to get worse at most by a factor of two for harder spectral indices ($E^{-2}$+cutoff scenario).
%This allows to directly compare the neutrino flux limits from Blazars with the diffuse neutrino flux and constrain the maximal contribution of Blazars given optimistic (weighting scheme 1) or conservative (weighting scheme 2) assumptions about the intrinsic $\gamma-\nu$ correlation. 
Furthermore, one can go beyond the 2LAC sources, since extrapolation of the gamma-ray luminosity function shows that the 2LAC Blazars (FSRQ's) already resolve more than \SI{60}{\percent} \cite{Inoue2011} (\SI{70}{\percent}\cite{Ajello2012}) of the EGB contribution from all Blazars (FRSQ's) in the observable universe. Thus, under the condition that the correlation between the $\gamma$ and $\nu$ energy flux holds (i.e. hadronic processes dominate the gamma-ray emission), one could further extend the "$\gamma$-Lum. weighting" limits to apply to the total corresponding population by dividing by $0.6$ ($0.7$).

\bibliography{ricap_14_arxiv_v1}

%\begin{thebibliography}{}
%Journal Author, Journal \textbf{Volume}, page numbers (year)
%Book Author, \textit{Book title} (Publisher, place, year) page numbers
% etc
%\end{thebibliography}

\end{document}